\documentclass[a4paper]{jpconf}
\usepackage{graphicx}
\begin{document}
\title{Orbital degeneracy and Mott transition in \\ Mo pyrochlore oxides}

\author{Yukitoshi Motome and Nobuo Furukawa$^{A,B}$}

\address{
Department of Applied Physics, University of Tokyo, %Tokyo 113-8656, 
Japan \\
$^A$Department of Physics and Mathematics, Aoyama Gakuin University, %Kanagawa 229-8558, 
Japan \\
$^B$Multiferroics Project, ERATO, Japan Science and Technology Agency (JST), c/o Department of Applied Physics, University of Tokyo, %Tokyo 113-8656,
 Japan}

\ead{motome@ap.t.u-tokyo.ac.jp}

\begin{abstract}
We present our theoretical results on an effective two-band double-exchange model 
on a pyrochlore lattice for understanding intricate phase competition in Mo pyrochlore oxides. 
The model includes the twofold degeneracy of $e_g'$ orbitals under trigonal field splitting, 
the interorbital Coulomb repulsion, the Hund's-rule coupling between 
itinerant $e_g'$ electrons and localized $a_{1g}$ spins, and 
the superexchange antiferromagnetic interaction between the $a_{1g}$ spins. 
By Monte Carlo simulation with treating the Coulomb repulsion 
at an unrestricted-type mean-field level, 
we obtain the low-temperature phase diagram as functions of the Coulomb repulsion and 
the superexchange interaction. 
The results include four dominant phases with characteristic spin and orbital orders 
and the metal-insulator transitions among them. 
The insulating region is characterized by a `ferro'-type orbital ordering 
of the $e_g'$ orbitals along the local $\langle 111 \rangle$ axis, irrespective of the spin ordering. 
\end{abstract}

\section{Introduction}
Mo pyrochlore oxides, whose chemical formula is generally given by $R_2$Mo$_2$O$_7$ 
($R$ is rare earth), are known to exhibit interesting phase competition and related phenomena. 
By chemical substitution of $R$ cations, 
the system changes from a ferromagnetic metal (FM) to a spin-glass (SG) insulator
systematically depending on the ionic radius of $R$ cation~\cite{Greedan1987}. 
On the other hand, in applied external pressure, 
FM collapses and turns into a peculiar paramagnetic metal (PM) 
with showing an intermediate SG metallic state, 
whereas the SG insulator remains rather robust against pressure~\cite{Iguchi2009}. 
This variety of phases and competition among them offer 
an opportunity to systematically study many fascinating phenomena in strongly correlated electron systems, 
such as metal-insulator transition and anomalous transport properties~\cite{Gardner2010}. 

The basic lattice structure of $R_2$Mo$_2$O$_7$ is composed of two intervening pyrochlore lattices: 
One is formed by Mo cations and the other is by $R$. 
The pyrochlore lattice is a three-dimensional network of corner-sharing tetrahedra, 
which is known to possess strong geometrical frustration. 
Mo cation is nominally tetravalent and has two electrons on average. 
The crystal field from trigonal distortion of MoO$_6$ octahedra 
splits three-fold $t_{2g}$ levels of Mo $4d$ electrons into $a_{1g}$ singlet and $e_g'$ doublet. 
The trigonal axis is along the direction toward the center of each Mo tetrahedron, 
which is called the local $\langle 111 \rangle$ axis. 
With aligning their spins parallel according to the Hund's rule, 
one of two electrons occupies the lower $a_{1g}$ level, and 
the other comes in the doubly-degenerate $e_g'$ levels. 
(See, for example, Fig.~1 in Ref.~\cite{Iguchi2009}.) 
Hence spin, charge, and orbital degrees of freedom are all active in these compounds, and 
the frustration of the lattice structure promotes the competition among them 
by preventing a simple ordering. 
In addition, the coupling between Mo $4d$ electrons and $R$-site rare-earth moments 
brings about further complications. 
In the following, however, we will focus on the phenomena 
commonly observed regardless of magnetic or nonmagnetic $R$ cations, 
such as the phase competition mentioned above, and neglect the $R$ sites: 
We will concentrate on the Mo pyrochlore network alone and 
the intrinsic physics from Mo $4d$ electrons in this paper. 

Several theoretical studies have been done for understanding the phase competition 
in the Mo pyrochlore oxides. 
A pioneering work was done for the chemical substitution at ambient pressure 
by Solovyev by using the LDA+$U$ calculation~\cite{Solovyev2003}. 
The calculated electronic structure indicates that 
the lower $a_{1g}$ level has rather localized nature with a narrow band, 
while the $e_g'$ levels form a relatively wide band and bears itinerant nature. 
The observation suggests that the system can be viewed 
as a double-exchange (DE) system, in which localized $a_{1g}$ spins 
are coupled with itinerant $e_g'$ electrons via the ferromagnetic Hund's-rule coupling. 
This explains well the emergence of FM 
in the compounds with large ionic radius of $R$, such as $R$=Nd and Sm. 
The calculation while changing the value of $U$ shows that 
the localization from FM to SG insulator by the substitution is basically understood 
by the correlation effect (or the bandwidth control). 
These calculations, however, were performed with assuming 
a simple collinear ferro or antiferromagnetic (AF) ordering; 
hence, the detailed nature of the frustrated insulating phase, 
such as the SG behavior and the orbital state, was not clarified. 

Recently, the authors have studied the phase competition under applied pressure~\cite{Motome2010A,Motome2010B,Motome2010C}.  
The results show that the change from FM, SG metal, to PM is well reproduced 
by Monte Carlo (MC) calculations for an extended DE model 
including the superexchange (SE) AF coupling between localized spins. 
In particular, the peculiar diffusive nature of PM is explained 
by the strongly incoherent state originating from competition 
between the DE ferromagnetic interaction and 
the SE AF interaction~\cite{Motome2010A}. 
The SG metallic state is also accounted by an instability toward phase separation 
related to the competition~\cite{Motome2010C}. 
The agreement implies that the physical pressure causes a different effect 
from the chemical pressure (chemical substitution), 
as implied in the experimental report~\cite{Iguchi2009}: 
The former tends to increase the SE AF coupling between the $a_{1g}$ spins dominantly, 
while the latter mainly modifies the bandwidth of itinerant $e_g'$ electrons. 
MC calculations so far were performed for a single-band model 
with neglecting the Coulomb repulsion between itinerant electrons; 
for comprehensive understanding of the phase competition in the Mo pyrochlore oxides, 
it is necessary to extend the analysis to incorporate 
the electron correlation as well as the $e_g'$ orbital degree of freedom. 

In this contribution, we investigate the effect of electron correlation and 
orbital degeneracy on the phase competition in the extended DE model. 
The model explicitly includes the twofold orbital degeneracy 
and the Coulomb repulsion between itinerant electrons. 
Thermodynamic properties of the model are studied by MC calculations, 
in which the electron correlation is handled by an unrestricted-type Hartree approximation. 
The results indicate that, in addition to the phase transition from FM to PM 
previously obtained in the single-band model, a metal-insulator transition is caused by 
the electronic correlation accompanied by a `ferro'-type orbital ordering. 
We discuss the phase diagram in relation to the distinct behaviors in experiments 
for the chemical substitution and the applied pressure in the Mo pyrochlore oxides. 

\if0{
The paper is organized as follows. 
In Sec.~2, we introduce our model and numerical method. 
MC results are shown in Sec.~3. Schematic phase diagram at low temperatures is deduced. 
In Sec.~4, we discuss the results in comparison with experiments 
with some concluding remarks. 
}\fi

\section{Model and method}
As an extension of the previous studies~\cite{Motome2010A,Motome2010B,Motome2010C}, 
we here consider the DE model with twofold orbital degeneracy. 
The Hamiltonian is given by
\begin{eqnarray}
{\cal H} &=& \sum_{\langle ij \rangle} \sum_{\alpha\beta} \sum_{\sigma} 
t_{\alpha\beta} (c_{i\alpha\sigma}^\dagger c_{j\beta\sigma} + {\rm H.c.}) 
+ \frac12 \sum_{i} \sum_{\alpha\beta\alpha'\beta'} \sum_{\sigma\sigma'} 
U_{\alpha\beta\alpha'\beta'} c_{i\alpha\sigma}^\dagger c_{i\beta\sigma'}^\dagger 
c_{i\beta'\sigma'} c_{i\alpha'\sigma} 
\nonumber \\
&&- J_{\rm H} \sum_i \sum_{\alpha} c_{i\alpha\sigma}^\dagger 
\vec{\sigma}_{\sigma\sigma'} c_{i\alpha\sigma'} \cdot \vec{S}_i 
+ J_{\rm AF} \sum_{\langle ij \rangle} \vec{S}_i \cdot \vec{S}_j, 
\label{eq:H_full}
\end{eqnarray}
where $c_{i\alpha\sigma}^\dagger$ ($c_{i\alpha\sigma}$) creates (annihilates) 
an electron with orbital $\alpha$ and spin $\sigma$ at site $i$, 
$\vec{\sigma}$ denotes the Pauli matrix, and 
$\vec{S}_i$ represents the localized spin at site $i$. 
The first term describes the electron hopping between nearest-neighbor sites 
$\langle ij \rangle$ on the pyrochlore lattice; 
the transfer integrals depend on the orbital indices $\alpha$ and $\beta$ 
which represent two $e_g'$ orbitals. 
The second term denotes the onsite Coulomb interactions between the $e_g'$ electrons, 
including the intra and interorbital repulsions, the Hund's-rule coupling, and the pair hopping. 
The third term denotes the Hund's-rule coupling between itinerant electrons and 
localized spins $\vec{S}_i$. 
The last term is the SE AF coupling between the neighboring localized spins. 
A similar model has been studied on the square or cubic lattices 
in the context of colossal magnetiresistive (CMR) manganites~\cite{Dagotto2001}. 

Our purpose is to clarify thermodynamics of the model given by Eq.~(\ref{eq:H_full}) 
and to elucidate the competing phases including correlated disordered states, 
such as the diffusive PM or the Mott insulating state. 
For this purpose, some sophisticated method 
beyond a simple mean-field approximation is necessary. 
The model (\ref{eq:H_full}), however, is very complicated and 
difficult to handle beyond a simple mean-field approximation. 
Hence, we introduce several simplifications as follows. 
First we take the limit of large $J_{\rm H}$. %, 
%as in the single-band model previously studied~\cite{Motome2010A,Motome2010B,Motome2010C}. 
In this limit, as often considered in the study of CMR, 
spins of itinerant electrons are aligned parallel to the localized spin at each site, 
and the transfer integrals are modulated by the relative angle 
between the neighboring localized spins. 
At the same time, among may contributions in the second term in Eq.~(\ref{eq:H_full}), 
only the interorbital interaction between the same spin electrons remains relevant. 
Furthermore, we consider the localized spins $\vec{S}_i$ as classical vectors 
with the renormalized length $|\vec{S}_i| = 1$. 
The simplifications were shown to give reasonable results in the study of 
a single-band model for the phase competition under pressure 
in the wide-bandwidth compounds~\cite{Motome2010A,Motome2010B,Motome2010C}.
Consequently, by taking the local spin axis along the direction of localized spin at each site, 
the model can be described by spinless fermions in the form 
\begin{equation}
{\cal H} = \sum_{\langle ij \rangle} \sum_{\alpha\beta} \tilde{t}_{i\alpha j\beta} 
(\tilde{c}_{i\alpha}^\dagger \tilde{c}_{j\beta} + {\rm H.c.}) 
+ \frac{\tilde{U}}{2} \sum_i \sum_{\alpha\neq\beta} \tilde{n}_{i\alpha} \tilde{n}_{i\beta} 
+  J_{\rm AF} \sum_{\langle ij \rangle} \vec{S}_i \cdot \vec{S}_j, 
\label{eq:H}
\end{equation}
where $\tilde{c}_{i\alpha}$ is the spinless fermion operator and  
$\tilde{n}_{i\alpha} = \tilde{c}_{i\alpha}^\dagger \tilde{c}_{i\alpha}$. %, and 
Here, $\tilde{U}$ is considered as an effective interorbital interaction, 
which controls the double occupancy in the $e_g'$ orbital with the same spin electrons 
under the strong Hund's-rule coupling; 
its origin is traced back to
the $e_g'$ interorbital Coulomb repulsion and the Hund's-rule couplings 
between $e_g'$-$e_g'$ and $e_g'$-$a_{1g}$ orbitals in Eq.~(\ref{eq:H_full}). 
Hereafter, we consider the filling with one electron per site on average. 

In Eq.~(\ref{eq:H}), the effective transfer integral is given by 
\begin{equation}
\tilde{t}_{i\alpha j\beta} = t_{\alpha\beta} 
\Big[ \cos \frac{\theta_i}{2} \cos \frac{\theta_j}{2} + \sin \frac{\theta_i}{2} \sin \frac{\theta_j}{2} 
\exp\{-i(\phi_i - \phi_j)\} 
\Big], 
\end{equation}
where $(\theta_i,\phi_i)$ denotes the angle of $\vec{S}_i$. 
To estimate the value of $t_{\alpha\beta}$, 
it is necessary to take account of the anisotropy of the $e_g'$ orbitals and 
the relative angle of Mo-O-Mo bond. 
Following the calculation in Ref.~\cite{Ichikawa2005}, 
we estimate the orbital diagonal and off-diagonal elements: 
it turns out that the ratio between them depends solely on the Mo-O-Mo bond angle $\delta$ as 
\begin{equation}
\frac{t_{\alpha \neq \beta}}{t_{\alpha=\beta}} = \frac{3-\cos\delta}{3+\cos\delta}.
\end{equation}
Thus, the orbital off-diagonal element is larger than the diagonal one for $\delta > 90^\circ$. 
In Mo pyrochlore oxides, $\delta$ is estimated around $130^\circ$ 
at ambient pressure~\cite{Moritomo2001}. 
In the following calculations, for simplicity, we consider only 
the dominant off-diagonal element and 
fix $t_{\alpha \neq \beta} = -1$ and $t_{\alpha=\beta}=0$, 
although the transfer integrals might change depending on both chemical and physical pressure. 

We study the model (\ref{eq:H}) by MC simulation in a similar way as in the previous studies 
for the single-band model~\cite{Motome2010A,Motome2010B,Motome2010C}.
To facilitate the MC calculations without negative sign problem, 
we introduce an unresticted-type mean-field decoupling 
to the Coulomb interaction in the second term in Eq.~(\ref{eq:H}). 
Here, as the simplest approximation, we employ the Hartree approximation; 
$\tilde{n}_{i\alpha} \tilde{n}_{i\beta} \to 
\langle \tilde{n}_{i\alpha} \rangle \tilde{n}_{i\beta} 
+ \tilde{n}_{i\alpha} \langle \tilde{n}_{i\beta} \rangle
- \langle \tilde{n}_{i\alpha} \rangle \langle \tilde{n}_{i\beta} \rangle$. 
The MC sampling is performed to update the configurations of localized spins $\{ \vec{S}_i \}$. 
In each MC step, the mean fields $\{ \langle n_{i\alpha} \rangle \}$ 
are recalculated by using the eigenvalues and eigenvectors 
obtained from the exact diagonalization of the Hamiltonian (\ref{eq:H}). 
We confirmed the convergence of the mean fields within the statistical errorbars 
after thermalization. 
We also employ the averaging over the twisted boundary conditions 
to suppress the finite size effect~\cite{Motome2010B}. 
In the following, we show the result for 16 site system with 
averaging over $8^3$ different twisted boundary conditions. 
Typically, we perform 2000 MC steps for measurement after 1000 MC steps for thermalization.

\section{Result}  
\begin{figure}[h]
\begin{minipage}{38pc}
\includegraphics[width=36pc]{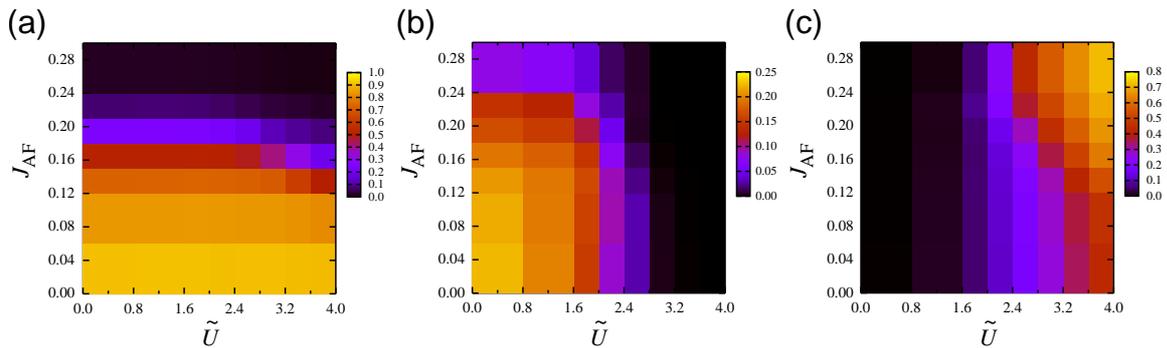}
\caption{\label{fig1}
MC results for the model given by Eq.~(\ref{eq:H}) at $\langle \tilde{n} \rangle =1$ and $T=0.04$: 
(a) square of the magnetization $m^2$ [Eq.~(\ref{eq:m^2})], 
(b) charge compressibility $\chi_{\rm c}$ [Eq.~(\ref{eq:chi_c})], and 
(c) $z$ component of orbital polarization $\tau^z$ [Eq.~(\ref{eq:tau^z})]. 
The results are obtained for 16 site cluster by taking average over twisted 
boundary conditions with $8^3$ different fluxes. 
}
\end{minipage}
\end{figure}

Figure~\ref{fig1} shows the MC results for the model (\ref{eq:H}) 
at a low temperature $T=0.04$ while changing 
the Coulomb repulsion $\tilde{U}$ and the SE AF coupling $J_{\rm AF}$. 
We compute the square of magnetization $m^2$, 
the charge compressibility $\chi_{\rm c}$, and 
the $z$ component of orbital polarization $\tau^z$, by 
\begin{eqnarray}
&& m^2 = \Big\langle \Big( \frac{1}{N_{\rm s}} \sum_i \vec{S}_i \Big)^2 \Big\rangle, 
\label{eq:m^2}
\\
&& \chi_{\rm c} = \frac{\partial \langle \tilde{n} \rangle}{\partial \mu} 
= \frac{\partial}{\partial \mu} 
\Big\langle \frac{1}{N_{\rm s}} \sum_{i\alpha} \tilde{n}_{i\alpha} \Big\rangle, 
\label{eq:chi_c}
\\
&& \tau^z = \Big\langle \Big|\frac{1}{N_{\rm s}} 
\sum_i (\tilde{n}_{i\alpha} - \tilde{n}_{i\beta})\Big| \Big\rangle, 
\label{eq:tau^z}
\end{eqnarray}
respectively. 
Here the bracket $\langle \cdots \rangle$ denotes the thermal average, 
$N_{\rm s}$ is the number of sites, 
$\mu$ is the chemical potential, and $\tilde{n}$ is the electron density. 

As shown in Fig.~\ref{fig1}(a), in the small $J_{\rm AF}$ region, 
the magnetization is large, indicating that the system is ferromagnetic. 
The ferromagnetism collapses as $J_{\rm AF}$ increases. 
On the other hand, the magnetism is relatively insensitive to $\tilde{U}$. 
For small $\tilde{U}$, the charge compressibility $\chi_{\rm c}$ is finite 
as shown in Fig.~\ref{fig1}(b), indicating that the system is metallic. 
Hence, the system is FM for small $J_{\rm AF}$, 
while it turns into PM as $J_{\rm AF}$ increases in the small $\tilde{U}$ region. 
This behavior is qualitatively the same as observed in the single-band model~\cite{Motome2010A}. 
In the single-band case, phase separation appears between the two phases~\cite{Motome2010C}; 
we expect a similar behavior in the present two-band case, 
but it is not clearly seen in the present calculation 
presumably because of the small size cluster. 

As $\tilde{U}$ increases, $\chi_{\rm c}$ decreases rapidly to zero, as shown in Fig.~\ref{fig1}(b). 
This signals the metal-to-insulator transition. 
The insulating state is characterized by an orbital ordering. 
Figure~\ref{fig1}(c) shows that the orbital polarization becomes nonzero 
almost simultaneously with the disappearance of the charge compressibility. 
This means that the insulating state shows a `ferro'-type orbital order 
by choosing one of two $e_g'$ orbitals. 
The reason for the `ferro'-type orbital order is understood 
from the strong coupling picture as follows. 
In the large $\tilde{U}$ limit, the second order perturbation in the transfer integrals 
leads to the so-called Kugel-Khomskii-type spin-orbital exchange couplings~\cite{Kugel1973}. 
In the present model, as seen in Fig.~\ref{fig1}(a), 
the spin configuration is governed by the competition between DE and SE interactions, 
not by $\tilde{U}$. 
Then, the `ferro'-type orbital configuration is favored via the spin-orbital exchange couplings,  
irrespective of spin configurations, 
because the dominant hopping processes in the perturbation are 
the orbital off-diagonal ones by $t_{\alpha \neq \beta}$ in the present model. 
The situation is contrastive to the `myth' for the Kugel-Khomskii-type interactions, 
in which the system usually favors complementary spin-orbital configurations, 
such as spin-AF orbital-ferro or spin-ferro orbital-AF. 

It is worthy to note that the `ferro'-type orbital order is not a uniform orbital order. 
The trigonal distortion is along the local $\langle 111 \rangle$ axis at each site, 
and hence, the $e_g'$ orbital polarization takes place along the site-dependent local axis. 
In other words, the `ferro'-type orbital order is a $\vec{q}=0$ four-sublattice order. 
Thus it does not break the cubic symmetry of the pyrochlore lattice structure. 

\begin{figure}[h]
\includegraphics[width=12.6pc]{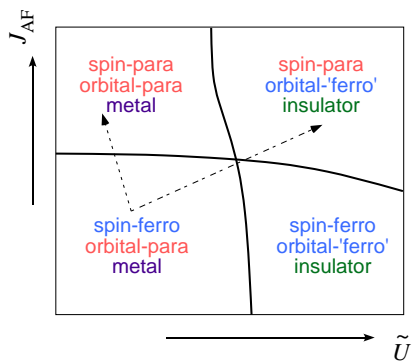}\hspace{2pc}%
\begin{minipage}[b]{23pc}\caption{\label{fig2}
Schematic phase diagram at low temperatures suggested from the results in Fig.~\ref{fig1}. 
The dotted (dot-dashed) arrow in the figure indicates a possible parameter change 
while changing the physical (chemical) pressure in the Mo pyrochlore oxides. 
See the text for details. 
}
\end{minipage}
\end{figure}

Summarizing the results, we obtain four dominant phases at low temperatures 
in the plane of $\tilde{U}$ and $J_{\rm AF}$, as schematically shown in Fig.~\ref{fig2}. 
The spin-ferro orbital-para metal in the small $\tilde{U}$ and small $J_{\rm AF}$ region 
is stabilized by the DE mechanism. 
The spin-para orbital-para metal in the small $\tilde{U}$ and large $J_{\rm AF}$ region 
is induced by the competition 
between the DE ferromagnetic interaction and the SE AF coupling 
under the geometrical frustration. 
In these two phases, the orbital degree of freedom is less important, 
and hence, the physics is common to the single-band case~\cite{Motome2010A,Motome2010B, Motome2010C}. 
On the other hand, the spin-ferro and orbital-`ferro' insulator 
in the large $\tilde{U}$ and small $J_{\rm AF}$ region is a correlation-driven insulator 
with development of the orbital ordering. 
The spin-para orbital-`ferro' insulator is driven by the increase of $J_{\rm AF}$ and 
the spin frustration. 

\section{Discussion and concluding remark}
In Mo pyrochlore oxides, as mentioned in Sec.~1, 
the chemical substitution leads to the transition from FM to SG insulator, 
while the pressure induces the transition from FM to PM. 
The latter is qualitatively explained by the change of $J_{\rm AF}$ 
in the single-band model~\cite{Motome2010A}, and 
the present calculation for the extended two-band model also gives similar results 
in the small $\tilde{U}$ regime; 
the parameter change is, e.g., schematically shown by the dotted arrow in Fig.~\ref{fig2}. 
On the other hand, the metal-insulator transition by the chemical substitution 
cannot be understood by the single-band physics. 
From our results, we deduce that the transition 
from FM to SG insulator is described by the transition 
from the spin-ferro orbital-para metal to the  spin-para orbital-`ferro' insulator, 
as indicated by the dot-dashed arrow in Fig.~\ref{fig2}. 
In the spin-para orbital-`ferro' insulating state, 
the spin degree of freedom remains paramagnetic because of the strong frustration, 
but short-range AF correlations are expected by the dominant $J_{\rm AF}$. 
The remaining frustration might be released by remnant perturbations neglected in the present model. 
In fact, one of the author and his collaborators recently demonstrated that 
the frustrated spins have instability toward a SG state 
through the coupling to local lattice distortions in the presence of bond disorder, 
reasonably in accord  with the experiments in Mo pyrochlore oxides~\cite{ShinaokaPreprint}. 
Thus, our results provide a hint for understanding 
the nature of the frustrated Mott insulating state in the Mo pyrochlore oxides. 
From these observations, we consider that, despite several simplifications introduced, 
our effective two-band model given by Eq.~(\ref{eq:H}) can offer a simple starting point 
for comprehensively understanding the intricate physics $R_2$Mo$_2$O$_7$. 
Further analysis in larger system sizes is desired 
to elucidate the thermodynamics as well as transport properties. 

Why and how the two effects, the chemical substitution and the pressure, 
lead to such different behaviors remain unclear. 
In the present scenario, these effects are attributed to the different control of 
the Coulomb repulsion (or the bandwidth) and the SE AF coupling between $a_{1g}$ spins. 
The bandwidth is basically determined by the transfer integrals 
between the itinerant $e_g'$ orbitals, i.e., $t_{\alpha\beta}$ in our model, 
on the other hand, the SE AF coupling $J_{\rm AF}$ is approximately 
set by $t_{a_{1g}}^2/U_{a_{1g}}$, where $t_{a_{1g}}$ and $U_{a_{1g}}$ are 
the transfer integral between the $a_{1g}$ orbitals and 
the intraorbital Coulomb repulsion in the $a_{1g}$ orbital, respectively. 
Both $t_{\alpha\beta}$ and $t_{a_{1g}}$ are modified by the chemical or physical pressure 
via the change of the angle and/or bond length of Mo-O-Mo bonds. 
In order to clarify how they are modified and to distinguish the two effects, 
it is highly desired to study the microscopic change of the lattice parameters 
by, e.g., the detailed x-ray scattering experiment or the first-principles calculations 
with structural optimization. 

The authors thank Y. Tokura, S. Iguchi, and H. Shinaoka for fruitful discussions. 
This work was supported by Grant-in-Aids (Nos.~19052008 and 21340090), 
Global COE Program ``the Physical Sciences Frontier", and 
the Next Generation Super Computing Project, Nanoscience Program, MEXT, Japan.

\end{document}